# Measurement of the Interactions and Stability of MTDC Systems

Wanning Zheng, *IEEE Member*, Li Chai, *IEEE Member*

[1]***Abstract*− The small-signal stability of multi-terminal HVDC systems, which is related to the dynamic interactions among different VSCs through the coupling of DC and AC networks, has become one of the important issues for the safety and stable operation of modern power systems. On the other hand, the robust stability theory with ν-gap metric is an effective tool for the stability analysis and synthesis of uncertain feedback systems. In this paper, we combine it with the self-/en-stabilizing coefficients method to measure the relative stability and analyze the stability influence of different paths of interactions in an MTDC system. The stability index is defined to represent the stability margin with respect to different paths of interactions in an MTDC system. A method for calculating the range of uncertain parameters preserving the stability is presented based on the stability criterion. The influence of control parameters on robust stability through interactions among VSCs can be analyzed quantitatively. Extensive examples are given to demonstrate the application and the effectiveness of the proposed method.**

## NOMENCLATURE

| | |
|---|---|
| $P_{in}$, $P_{out}$ | Active power input and output of VSC |
| $Q$ | Reactive power output of VSC |
| $E$, $\theta$ | Internal voltage's magnitude and phase |
| $U_t$, $\theta_t$ | Terminal voltage magnitude and phase |
| $U_s$, $\theta_s$ | Infinite-bus voltage magnitude and phase |
| $\boldsymbol{I}$ | VSC current vector |
| $U_{dc}$ | DC voltage |
| $C$ | DC-capacitance |
| $X_f$, $C_f$ | VSC filter reactance and capacitance |
| $X_g$ | Transmission line reactance in AC system |
| $\mathrm{PI}_j = k_{pj}+k_{ij}/s$ | A generic PI controller (j=1, 2,…,6) |
| $R$, $L$ | Equivalent resistance and inductance of DC cables |
| $G_m(s)$, $D(s)$ | Transfer functions related to inertia and damping |
| $G_{in}(s)$ | Transfer functions between active power input and phase of internal voltage |
| $G_q(s)$ | Transfer functions related to reactive power control |
| Superscript: | |
| ~ | Parameters or transfer functions with uncertainties |
| Subscripts: | |
| *0* | Initial values in steady-state condition |

## I. INTRODUCTION

Multi-terminal high voltage direct current (MTDC) systems based on voltage source converters (VSCs) are widely used to integrate and transmit large-scale renewable generations [1]-[2]. It is important and challenging to study the stability property of an MTDC system since its dynamic behavior is related to the complex interactions among different VSCs through the coupling of AC and DC networks. Many researchers try to study the influence mechanism of interactions on the system stability, especially to measure the interactions and the stability of the system affected by controller parameters.

Participation factors and sensitivity analysis of modal analysis are widely used to analyze the interactions in MTDC systems [3]-[6]. However, it is reported lack of mechanism understanding on how interactions cause the instability [7]. Therefore, open-loop modal coupling method is proposed in [7]-[8] to reveal the mechanism of oscillations caused by the interactions. The results show when the modes of an open loop VSC subsystem and the modes of the rest part of the system are close to each other, the open-loop modal coupling occurs and would affect the stability of an MTDC system. In addition, based on the impedance model of the AC grid-connected VSCs, interactions are characterized by the participation factors and sensitivity under this method in frequency domain [9]. The influence of the inner and outer loop control parameters on the impedance characteristics of VSCs and the stability of the system are analyzed as well [10]-[14].

In order to present meaningful insights about interactions, such as which devices or interaction paths cause the oscillations and how control parameters affect the interactions and the system stability, [15] proposed an analytic quantification method for the interaction analysis of MTDC systems. The MTDC system is described as a closed-loop system composed by the self-stabilizing coefficient and the en-stabilizing coefficients at the feedback channels. Under this setup, interactions among VSCs can be quantified by analytic formulas.

However, less is known about the stability margin of the MTDC system and the ranges of uncertain parameters preserving the stability.

To solve this problem, the authors put forward an idea of combining robust control theory with the interaction quantification method in [15] together. As we know, robust

[1] This work was supported in part by the National Natural Science Foundation of China under Grant 62173259 and Grant 61625305.

The authors are with the Engineering Research Center of Metallurgical Automation and Measurement Technology, Wuhan University of Science and Technology, Wuhan 430081, China (e-mail: zhengwanning@wust.edu.cn; chaili@wust.edu.cn).

control theory is proposed to evaluate the anti-interference ability of uncertainties such as control parameters. *Mu* method and *H*∞ control are widely used to analyze the stability and design the controllers in power systems with VSCs [16]-[19]. However, those methods will result in a dynamic controller whose order is very high (usually the same as that of the system). This feature makes it more difficult to know about the stability margin and how controllers affect the system stability through interactions. Actually, engineers prefer to optimize parameters in existing simple control framework to achieve better stability margin through theoretical analysis, rather than implementing the high order controller with new structure.

The robust stability theorem developed on ν-gap metric is an elegant and very effective tool for the analysis and synthesis of uncertain feedback systems [20]. Roughly speaking, the ν-gap metric is proposed to measure the difference of two systems when a feedback controller is applied. Unfortunately, it is familiar only to experts on the robust control theory, not well-known even in the general control community, leave alone researchers outside the control community.

This paper applies the ν-gap metric to the closed-loop description of MTDC systems obtained by the interaction quantification method in [15]. The main contributions are as follows: (i) A stability index is proposed to reflect the stability margin of an MTDC system with respect to different paths of interactions. (ii) A method for calculating the ranges of uncertain parameters preserving the stability is presented based on the stability criterion composed of ν-gap metric and the stability index. The stability sensitiveness can also be analyzed by this method based on the variation of the ν-gap between certain and uncertain systems with respect to the variation of a particular parameter or interaction. To the best of our knowledge, this is the first time that the ν-gap metric is applied to the stability analysis of power systems. The results in this paper can explain the influence of control parameters on robust stability through interactions among VSCs quantitatively, and provide new ideas for multi-equipment-based stability control design.

The rest of this paper is organized as follows. Section II presents the background knowledge, including typical control and small signal model of an MTDC system, analytic quantification of interactions and ν-gap metric. In Section III, a method to measure robust stability and interactions of an MTDC system is proposed. Case study is proposed in Section IV, and conclusions are drawn in Section V.

## II. Background Knowledge

In this section, we present some background knowledge on the self-/en-stabilizing coefficient method of analyzing interactions among VSCs, and the robust stability result based on ν-gap metric. For more detail, please refer to [15, 23-24], [20] [Chapter 8, 21]. For easy understanding and compact representation, we change some notations, which are slightly different to the original literature.

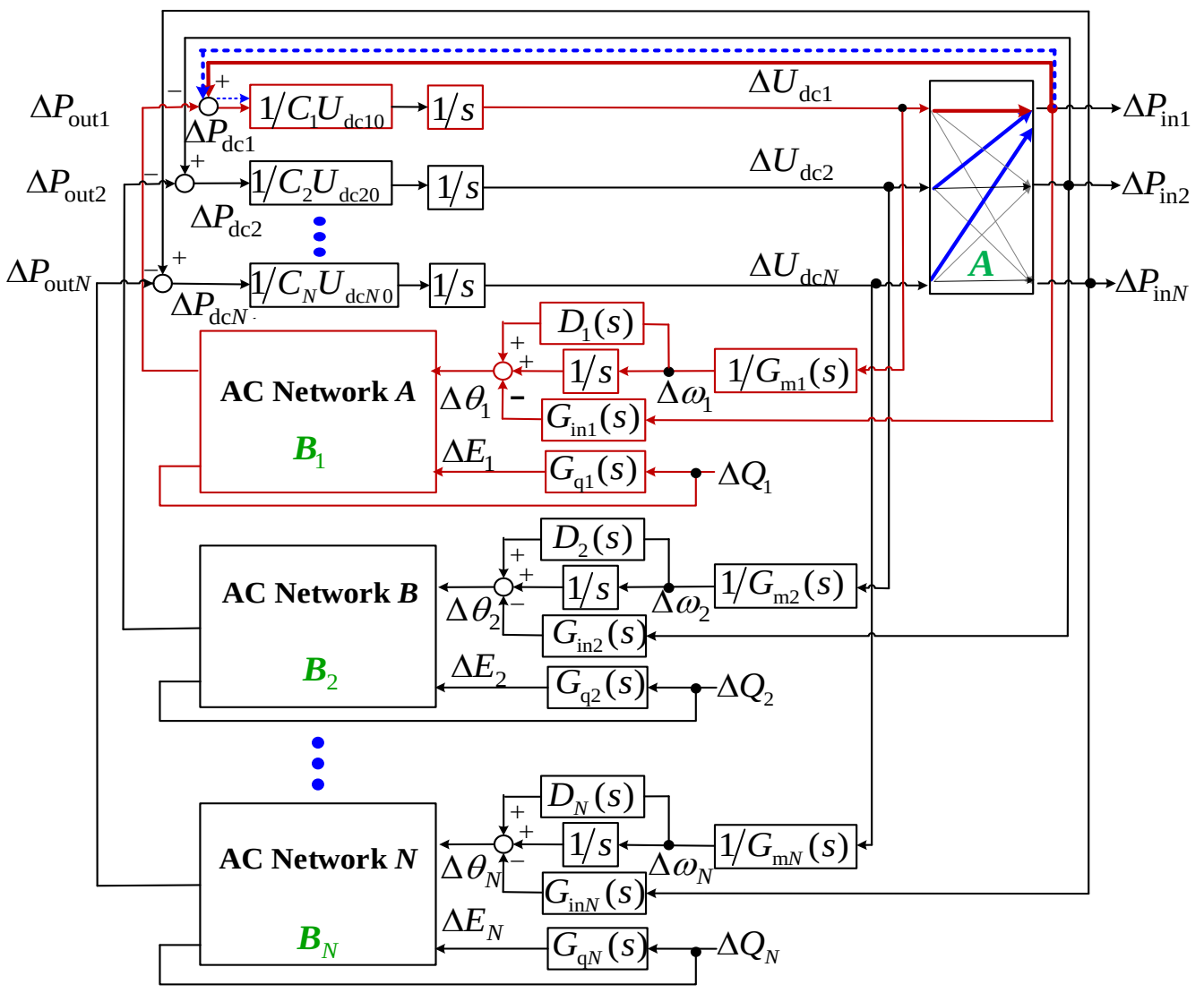

Fig. 1. The system model of an MTDC system based on the motion equation concept.

### *A. Typical control of an MTDC system*

I-U DC voltage-droop control cascaded with vector control and phase-locked loop (PLL) is typically designed for two-level VSCs in an MTDC system. To balance the active power flow in such systems, I-U DC voltage-droop control generates the reference value of DC voltage according to the actual DC current in each VSC. Then, the vector control manipulates the DC voltage and reactive power of each VSC in a decoupled manner in PLL-synchronized reference frame.

### *B. Small signal model of an MTDC system in DC voltage control timescale*

On the basis of the model proposed in [23], consider the small signal model of an MTDC system with *N* VSCs in DC voltage control timescale as shown by Fig.1 [15]. For the *i*-th VSC, $i$=1, 2, … $N$, let $\Delta P_{\mathrm{in}i}$, $\Delta P_{\mathrm{out}i}$, and $\Delta Q_i$ be the active power input from DC networks, the active power output and reactive power from AC networks, respectively, and let $\Delta U_{\mathrm{dc}i}$ be the DC voltage to DC network, $\Delta\theta_i$ and $\Delta E_i$ the phase and amplitude of internal voltage to AC network. Denote

$$\Delta \boldsymbol{P}_{\mathrm{in}}=[\Delta P_{\mathrm{in}1}\ \Delta P_{\mathrm{in}2}\ldots\Delta P_{\mathrm{in}N}\ ]^{\mathrm{T}},$$
$$\Delta \boldsymbol{U}_{\mathrm{dc}}=[\Delta U_{\mathrm{dc}1}\ \Delta U_{\mathrm{dc}2}\ldots\Delta U_{\mathrm{dc}N}]^{\mathrm{T}}.$$

Then the system can be described as follows

$$\Delta \boldsymbol{P}_{\mathrm{in}}=\left[A_{ij}\right]\Delta \boldsymbol{U}_{\mathrm{dc}} \tag{1}$$

$$\begin{bmatrix}\Delta P_{\mathrm{out}i}\\ \Delta Q_i\end{bmatrix}=\boldsymbol{B}_i\begin{bmatrix}\Delta\theta_i\\ \Delta E_i\end{bmatrix}=\begin{bmatrix}K_{\mathrm{P\theta}i} & K_{\mathrm{PE}i}\\ K_{\mathrm{Q\theta}i} & K_{\mathrm{QE}i}\end{bmatrix}\begin{bmatrix}\Delta\theta_i\\ \Delta E_i\end{bmatrix} \tag{2}$$

where $A_{\mathrm{ij}}=\begin{cases}I_{\mathrm{dc}i0}+y_{ii}U_{\mathrm{dc}i0} & i=j\\ y_{ij}U_{\mathrm{dc}i0} & i\neq j\end{cases}$,

$$K_{\mathrm{P\theta}i}=\frac{E_{i0}U_{gi0}\cos(\theta_{i0}\text{-}\theta_{gi0})}{X_{\mathrm{f}i}+X_{gi}},K_{\mathrm{PE}i}=\frac{U_{gi0}\sin(\theta_{i0}\text{-}\theta_{gi0})}{X_{\mathrm{f}i}+X_{gi}},$$

$$K_{\mathrm{Q\theta}i}=\frac{E_{i0}U_{gi0}\sin(\theta_{i0}\text{-}\theta_{gi0})}{X_{\mathrm{f}i}+X_{gi}},K_{\mathrm{QE}i}=\frac{2E_{i0}-U_{gi0}\cos(\theta_{i0}\text{-}\theta_{gi0})}{X_{\mathrm{f}i}+X_{gi}}.$$

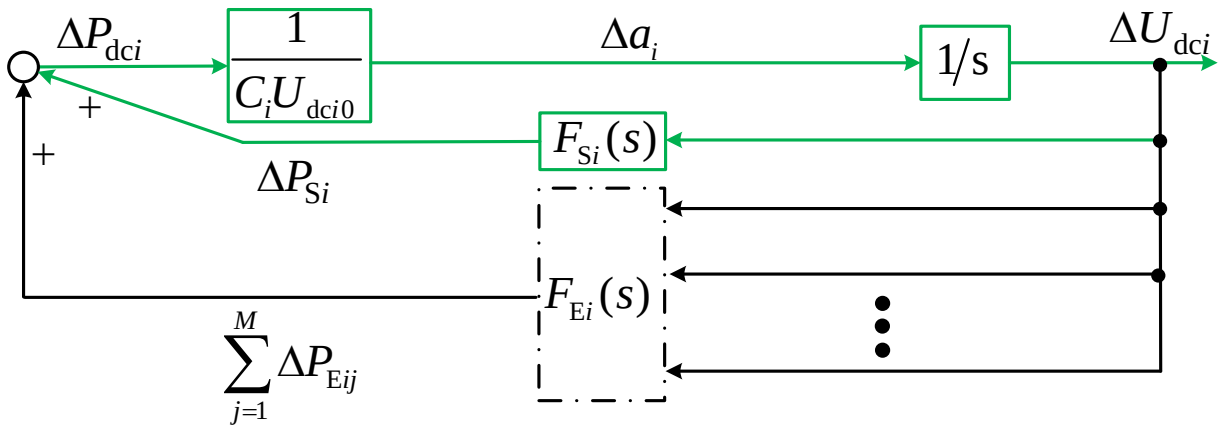

Fig. 2. An equivalent closed-loop MTDC system

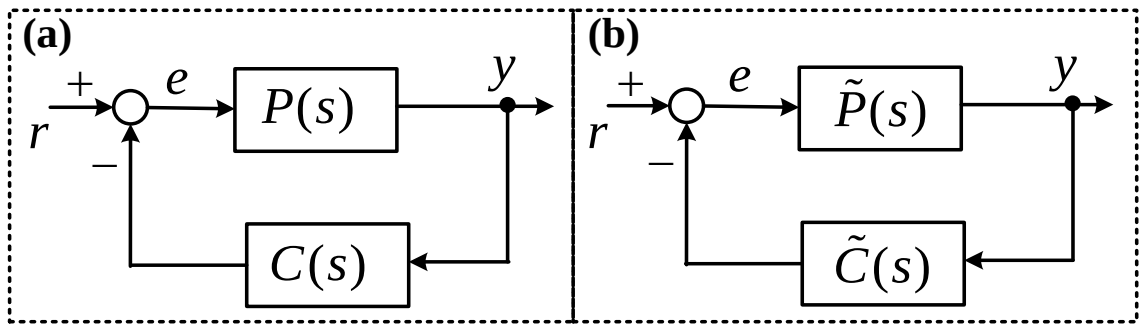

Fig. 3. A stable feedback system and an uncertain feedback system

*C. Analytic quantification of interactions among VSCs based on self-/en-stabilizing coefficients*

We can see from Fig. 1 and (1)-(2) that the dynamics of the $i$-th VSC are not only determined by its own characteristics, but also influenced by other VSCs' characteristics. We now introduce the self-/en-stabilizing coefficient method for the interaction analysis of MTDC systems proposed in [15].

The unbalanced active power across the DC capacitor of the $i$-th VSC can be written as

$$\begin{aligned}\Delta P_{\mathrm{dc}i} &= \Delta P_{\mathrm{in}i} - \Delta P_{\mathrm{out}i} \\ &= \left( A_{ii}\Delta U_{\mathrm{dc}i} + \sum_{j\neq i} A_{ij}\Delta U_{\mathrm{dc}j} \right) - \left( K_{\mathrm{P}\theta i}\Delta\theta_i + K_{\mathrm{PE}i}\Delta E_i \right)\end{aligned} \tag{3}$$

Denote $\Delta P_{\mathrm{S}i}$ and $\Delta P_{\mathrm{E}i}$ as the unbalanced active power proposed by VSC $i$ and by VSCs except $i$, that is

$$\Delta P_{\mathrm{S}i} = A_{ii}\Delta U_{\mathrm{dc}i} - K_{\mathrm{P}\theta i}\Delta\theta_i - K_{\mathrm{PE}i}\Delta E_i \tag{4}$$

$$\Delta P_{\mathrm{E}i} = \sum_{j\neq i} A_{ij}\Delta U_{\mathrm{dc}j} \tag{5}$$

The transfer function $F_{\mathrm{S}i}(s)=\Delta P_{\mathrm{S}i}(s)/\Delta U_{\mathrm{dc}i}$ is called the self-stabilizing coefficient, and the transfer function $F_{\mathrm{E}i}(s)=\Delta P_{\mathrm{E}i}(s)/\Delta U_{\mathrm{dc}i}$ is called the en-stabilizing coefficient. Then (3) can be expressed as

$$\Delta P_{\mathrm{dc}i} = F_{\mathrm{S}i}(s)\Delta U_{\mathrm{dc}i} + F_{\mathrm{E}i}(s)\Delta U_{\mathrm{dc}i} \tag{6}$$

As shown in [15], $F_{\mathrm{E}i}(s)$ can be further written as the sum of $M$ terms representing different paths according to the quantity of VSCs participating in the interactions and different effects of DC and AC networks, that is

$$F_{\mathrm{E}i}(s) = \sum_{j=1}^{M} F_{\mathrm{E}ij}(s) \tag{7}$$

Then, the system model in Fig. 1 can be converted into an equivalent closed-loop system as shown in Fig. 2.

*D. The ν-gap metric and robust stability margin*

The ν-gap metric between two LTI systems $G_1(s)$ and $G_2(s)$, denoted by $\nu[G_1(s), G_2(s)]$, is defined as follows.

$$\nu\left[G_1(s),G_2(s)\right] = \begin{cases} \max\limits_{\omega\in\mathbb{R}} \arcsin \dfrac{\left|G_1(j\omega)-G_2(j\omega)\right|}{\sqrt{1+\left|G_1(j\omega)\right|^2}\sqrt{1+\left|G_2(j\omega)\right|^2}} \\ \qquad\qquad\qquad if\ G_1(s)\sim G_2(s) \\ \dfrac{\pi}{2}, \qquad otherwise \end{cases} \tag{8}$$

where $G_1(s)\sim G_2(s)$ means that $G_1(s)$ and $G_2(s)$ are comparable. Generally, if $G_1(s)$ and $G_2(s)$ have the same number of zeros and poles in the left and right half planes, $G_1(s)$ and $G_2(s)$ are comparable. The rigor definitions of being comparable is presented in [21].

It is proved in [20] that ν-gap is a distance function. For a given system $G(s)$, we describe a set of uncertain systems around $G(s)$ as a ball with center $G(s)$ and radius $r$ measured by ν-gap:

$$\mathcal{B}_\nu\left[G(s),r\right] = \left\{\tilde{G}(s):\nu\left[G(s),\tilde{G}(s)\right]\leq r\right\} \tag{9}$$

The center $G(s)$ is called the nominal system.

Consider the standard feedback control system shown by Fig. 3(a), where $P(s)$ and $C(s)$ are the plant and controller respectively. Assume the system is stable, the robust stability margin of the feedback system is defined as

$$\begin{aligned} & b\left[P(s),C(s)\right] \\ &= \min_{\omega\in\mathbb{R}} \arcsin \frac{\left|P(j\omega)+C^{-1}(j\omega)\right|}{\sqrt{1+\left|P(j\omega)\right|^2}\sqrt{1+\left|C^{-1}(j\omega)\right|^2}}\end{aligned} \tag{10}$$

The robust stability margin is 0 if the system is unstable.

Now consider the feedback system depicted in Fig. 3(b), where $\tilde{P}(s)$ is the perturbed version of $P(s)$, and $\tilde{C}(s)$ is the perturbed version of $C(s)$. If the uncertain systems $\tilde{P}(s)$ and $\tilde{C}(s)$ are measured by ν-gap, we have the following result (Theorem 8.30 in [21]).

*Theorem 1*. Assume that the feedback system in Fig. 3(a) is stable, and the robust stability margin is $b[P(s), C(s)]$ defined by (10). Then the feedback system in Fig. 3(b) is stable for all $\tilde{P}(s) \in \mathcal{B}_\nu[P(s), r_\mathrm{P}]$ and $\tilde{C}(s) \in \mathcal{B}_\nu[C(s), r_\mathrm{C}]$ if and only if

$$r_\mathrm{P} + r_\mathrm{C} \leq b\left[P(s),C(s)\right] \tag{11}$$

Theorem 1 shows that the uncertain feedback system is stable if and only if the sum of ν-gap between the nominal systems and the uncertain systems is not greater than the robust stability margin of nominal feedback system. Generally speaking, the larger $b[P(s), C(s)]$, the better the relative stability of the nominal feedback system is.

## III. Measurement of Robust Stability and Interactions

In this section, we give the definition of the stability index of an MTDC system after transferring the closed-loop system in Fig. 2 to the standard feedback system as shown in Fig. 3. The stability criterion is obtained to address the problem of how uncertain parameters influence the stability of a particular VSC through different paths of interactions. A method to calculate the range of uncertain parameters preserving the stability is also proposed.

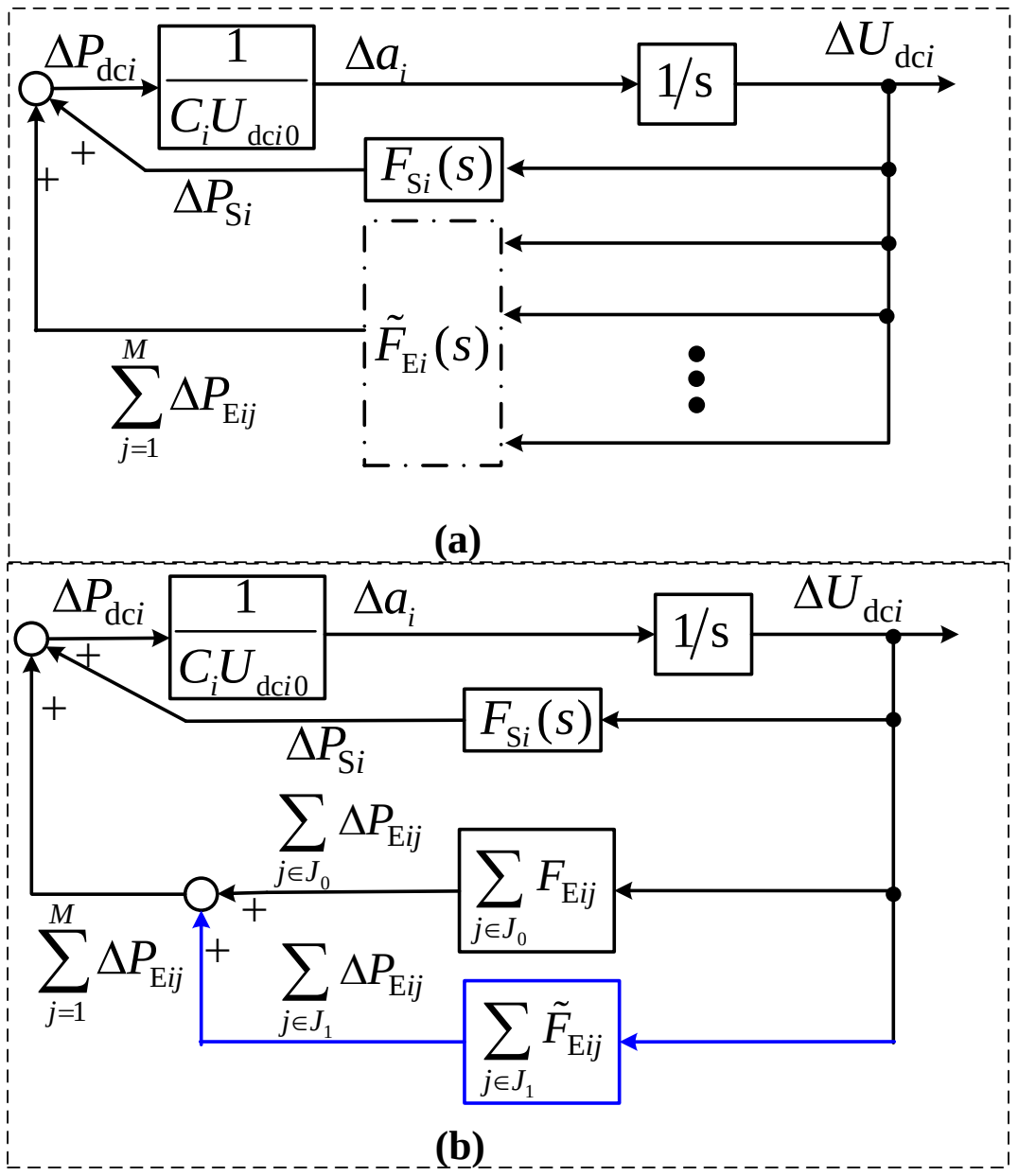

Fig. 4. An equivalent model of an MTDC system with uncertainties.

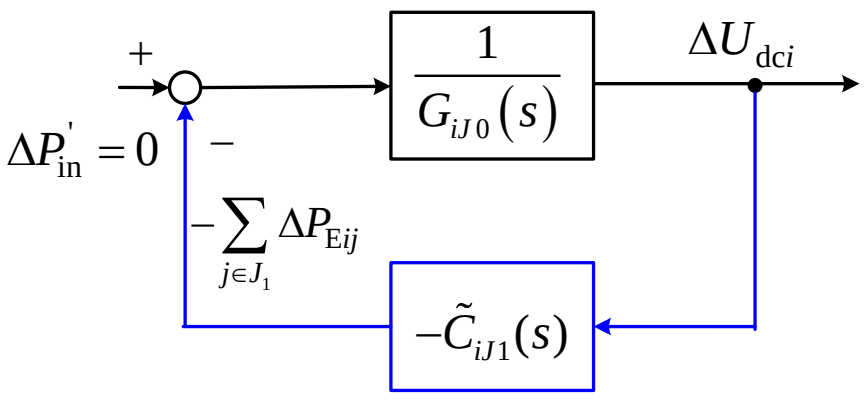

Fig. 5. An equivalent model for the application of ν-gap metric and stability index.

Consider the equivalent model of an MTDC system in Fig. 2 which reflects the interactions among VSCs. Assume that the feedforward channel and $F_{\mathrm{S}i}(s)$ of the $i$-th VSC are fixed. Then the equivalent model of an MTDC system with uncertainties can be depicted as Fig. 4 (a).

Recall that the en-stabilizing coefficients in (7) consists of $M$ terms representing different paths of interactions. Let $J_0 \subset \{1,\ldots,M\}$ be the set of paths without uncertainty, $J_1 \subset \{1,\ldots,M\}$ the set of paths with uncertain parameters. Obviously, $J_0 \cup J_1 = \{1,\ldots,M\}$. Define

$$C_{iJ1}(s) = \sum_{j\in J_1} F_{Eij}(s) \tag{12}$$

Denote $\tilde{C}_{iJ1}(s)$ as the uncertain counterpart of $C_{iJ1}(s)$. Then the model in Fig. 4 (a) can be depicted as Fig. 4 (b), which needs to be converted to the standard feedback channel as Fig. 3 so that Theorem 1 can be applied to analyze how much uncertainty there can be in the paths of $J_1$ so that the system is still stable provided other components fixed.

Denote

$$G_{iJ0} = sC_i U_{\mathrm{dc}i0} - F_{\mathrm{S}i}(s) - \sum_{j\in J_0} F_{\mathrm{E}ij}(s) \tag{13}$$

The part without uncertain parameters tagged in black lines in Fig. 4 (b) can be viewed as an open-loop transfer function $1/G_{iJ0}(s)$. The part with uncertainty $-\tilde{C}_{iJ1}(s)$ tagged in blue lines can be viewed as a feedback controller. Then the feedback system in Fig. 4 (b) can be converted to the equivalent standard feedback system with the plant $1/G_{iJ0}(s)$ and the controller $-\tilde{C}_{iJ1}(s)$, as shown in Fig. 5.

We are now ready to present the stability index of the MTDC system.

*Definition 1.* For the $i$-th VSC of an MTDC system shown in Fig. 5, the stability index with respect to the set of interactions $J_1$ is defined as

$$\varsigma_{iJ1} = \min_{\omega\in\mathbb{R}} \arcsin \frac{\left| j\omega C_i U_{\mathrm{dc}i0} - F_{\mathrm{S}i}(j\omega) - \sum_{j=1}^{M} F_{\mathrm{E}ij}(j\omega) \right|}{\sqrt{\left(1+|G_{iJ0}(j\omega)|^2\right)\left(1+|C_{iJ1}(j\omega)|^2\right)}} \tag{14}$$

where $J_1 \subset \{1,\ldots,M\}$, $C_{iJ1}(s)$ and $G_{iJ0}(s)$ are defined in (12) and (13), respectively.

With the help of ν-gap, we can measure the uncertainty of the interactions through paths $J_1$ quantitatively. In particular, assume that $\tilde{C}_{iJ1}(s) \in \mathcal{B}_\nu[C_{iJ1}(s), r_{iJ1}]$, that is, $\nu[\tilde{C}_{iJ1}(s), C_{iJ1}(s)] \le r_{iJ1}$ for $\tilde{C}_{iJ1}(s)$ corresponding to any uncertain parameters. Then we have the following results.

*Theorem 2.* Consider the stable MTDC system shown in Fig. 5. Let $\zeta_{iJ1}$ be the stable index with respect to $J_1$ defined by (14). Then the system is stable for all $\tilde{C}_{iJ1}(s) \in \mathcal{B}_\nu[C_{iJ1}(s), r_{iJ1}]$ if and only if

$$r_{iJ1} \le \xi_{iJ1} \tag{15}$$

*Proof.* By direct computation, we have

$$\begin{aligned}
&\frac{\left|\frac{1}{G_{iJ0}(j\omega)} - \frac{1}{C_{iJ1}(j\omega)}\right|}{\sqrt{1+\left|\frac{1}{G_{iJ0}(j\omega)}\right|^2}\sqrt{1+\left|\frac{1}{C_{iJ1}(j\omega)}\right|^2}} \\
&= \frac{|G_{iJ0}(j\omega) - C_{iJ1}(j\omega)|}{\sqrt{1+|G_{iJ0}(j\omega)|^2}\sqrt{1+|C_{iJ1}(j\omega)|^2}} \\
&= \frac{\left| j\omega C_i U_{\mathrm{dc}i0} - F_{\mathrm{S}i}(j\omega) - \sum_{j\in J_0} F_{Eij}(s) - \sum_{j\in J_1} F_{Eij}(s) \right|}{\sqrt{\left(1+|G_{iJ0}(j\omega)|^2\right)\left(1+|C_{iJ1}(j\omega)|^2\right)}} \\
&= \frac{\left| j\omega C_i U_{\mathrm{dc}i0} - F_{\mathrm{S}i}(j\omega) - \sum_{j=1}^{M} F_{Eij}(j\omega) \right|}{\sqrt{\left(1+|G_{iJ0}(j\omega)|^2\right)\left(1+|C_{iJ1}(j\omega)|^2\right)}}
\end{aligned} \tag{16}$$

It follows from (10) that

$$\begin{aligned}
&b\left[\frac{1}{G_{iJ0}(s)}, -C_{iJ1}(s)\right] \\
&= \min_{\omega\in\mathbb{R}} \arcsin \frac{\left|\frac{1}{G_{iJ0}(j\omega)} - \frac{1}{C_{iJ1}(j\omega)}\right|}{\sqrt{\left(1+\left|\frac{1}{G_{iJ0}(j\omega)}\right|^2\right)\left(1+\left|\frac{1}{C_{iJ1}(j\omega)}\right|^2\right)}}
\end{aligned} \tag{17}$$

Hence $\zeta_{iJ1}= b[1/G_{iJ0}(s), -C_{iJ1}(s)]$. Utilizing Theorem 1 with $r_P$=0, we obtain the results of Theorem 2. The proof is completed. ■

The stability index $\zeta_{iJ1}$ has significant meanings. On one hand, under the same feedforward and feedback channels, i.e. $G_{iJ0}(s)$ and $-\tilde{C}_{iJ1}(s)$, $\zeta_{iJ1}$ can reflect the overall stability of the system under different sets of parameters. The larger the index, the better the stability performance of the system. And one can quantitatively know how far a system is away from instability from $\zeta_{iJ1}$. On the other hand, under a particular set of parameters, the index can reflect the stability respect to different paths of interactions, viz. different $G_{iJ0}(s)$ and $-\tilde{C}_{iJ1}(s)$. The larger the index, the larger range of uncertainties can be tolerated through this paths of interactions.

For better use of Theorem 2, $r_{iJ1}$ can be calculated and simplified as follows. In this paper, $C_{iJ1}(s)$ and $\tilde{C}_{iJ1}(s)$ are all comparable. According to the definition of ν-gap metric, $r_{iJ1}$ can be written as

$$\begin{aligned} r_{iJ1} &= \max_{\omega\in\mathbb{R}} \arcsin \frac{\left|C_{iJ1}(j\omega)-\tilde{C}_{iJ1}(j\omega)\right|}{\sqrt{1+\left|C_{iJ1}(j\omega)\right|^2}\sqrt{1+\left|\tilde{C}_{iJ1}(j\omega)\right|^2}} \\ &= \max_{\omega\in\mathbb{R}} \arcsin \sqrt{\frac{\left|C_{iJ1}(j\omega)-\tilde{C}_{iJ1}(j\omega)\right|^2}{\left(1+\left|C_{iJ1}(j\omega)\right|^2\right)\left(1+\left|\tilde{C}_{iJ1}(j\omega)\right|^2\right)}} \end{aligned} \tag{18}$$

As we know, the squares of the absolute values in (18) can be expanded as

$$\begin{aligned} &\left|C_{iJ1}(j\omega)-\tilde{C}_{iJ1}(j\omega)\right|^2 \\ &=\left[C_{iJ1}(j\omega)-\tilde{C}_{iJ1}(j\omega)\right]\left[C_{iJ1}(-j\omega)-\tilde{C}_{iJ1}(-j\omega)\right], \\ &\left|C_{iJ1}(j\omega)\right|^2 = C_{iJ1}(-j\omega)C_{iJ1}(j\omega), \\ &\left|\tilde{C}_{iJ1}(j\omega)\right|^2 = \tilde{C}_{iJ1}(-j\omega)\tilde{C}_{iJ1}(j\omega) \end{aligned} \tag{19}$$

By substituting (19) into (18), $r_{iJ1}$ can be expressed and simplified as

$$r_{iJ1} = \max_{\omega\in\mathbb{R}} \arcsin\left\{1+\left|\frac{C_{iJ1}(j\omega)-\tilde{C}_{iJ1}(j\omega)}{1+C_{iJ1}(j\omega)\tilde{C}_{iJ1}(-j\omega)}\right|^{-2}\right\}^{-1/2} \tag{20}$$

As we know, a transfer function's infinite norm equals to the peak amplitude on the Bode magnitude plot of it. So $r_{iJ1}$ can be represented by infinite norm as

$$r_{iJ1} = \arcsin\left\{1+\left\|\frac{C_{iJ1}(s)-\tilde{C}_{iJ1}(s)}{1+C_{iJ1}(s)\tilde{C}_{iJ1}(-s)}\right\|_\infty^{-2}\right\}^{-1/2} \tag{21}$$

Utilizing Theorem 2, we can evaluate the stability of the system when the uncertain parameter is a specific value. Denote $x_i=\zeta_{iJ1}-r_{iJ1}$. The larger the $x_i$, the more stable the system is under this set of uncertain parameters. On the other hand, by calculating $r_{iJ1}$ under different uncertain parameters, the range of the uncertain parameters that satisfies the stability criterion can be obtained. When the uncertain parameter $c$ equals to a particular $c_0$, if $r_{iJ1}-\delta<\zeta_{iJ1}< r_{iJ1}+\delta$, in which $\delta$ is a very small positive real number, we use $c_0$ as the boundary of the uncertain parameter. If $r_{iJ1} \le \zeta_{iJ1}$ for any $c\le c_0$, then the range of the uncertain parameter is $c\le c_0$ from the perspective of stability.

It is worth noticing that the obtained range of parameters is not conservative. It means that for an MTDC system with a particular parameter $c=c_0$, if $r_{iJ1} \le \zeta_{iJ1}$, we can give the interval $[c_0-\beta_1, c_0+\beta_2]$ that the system remains stable for any $c$ within the interval. And for any $\varepsilon>0$, there exist $c_1$ such that i) the system is unstable and $r_{iJ1} > \zeta_{iJ1}$ with $c=c_1$ and other parameters fixed; ii) $c_0-\beta_1-c_1<\varepsilon$ or $c_1- c_0-\beta_2<\varepsilon$. In addition, we can analyze the stability sensitiveness of the MTDC system based on the variation of the ν-gap between certain and uncertain systems with respect to the variation of a particular parameter. The larger the variation of the ν-gap with different uncertain parameters, the more sensitive the parameters.

Therefore, the ν-gap metric and stability index are elegant and very effective tools for the analysis and synthesis of uncertain feedback systems. Case studies to evaluate the stability margin and the range of uncertain parameters of an MTDC system are presented in next section.

## IV. Case Study

Take a three-terminal HVDC system shown in Fig. 6 as the object, this section shows how to measure the stability margin produced by different paths of interactions and evaluate the ranges of uncertain parameters utilizing the method in Section III. In this three-terminal HVDC system, VSC *A*, *B* are rectifiers and *C* is an inverter. Since this paper mainly studies the method to measure the stability and interactions, AC grid in each terminal is assumed to be strong with the SCR larger than 3. The impact of reactive power branch to active power branch is omitted, which means $K_{PEi}$ and $K_{Q\theta i}$ equal to zero. Specific parameters are shown in Appendix *B*.

When focusing on VSC *A*, the equivalent closed-loop model in the form of Fig. 2 has six paths of interactions according to [15], which means $M$=6. $F_{S1}(s)$ represents the self-stabilizing coefficient of VSC *A*, and $F_{E11}(s)$~ $F_{E16}(s)$ represent different paths of en-stabilizing coefficients. $F_{E11}(s)$ represents single en-stabilizing coefficient from VSC *B* to *A* coupled only by DC network. $F_{E12}(s)$ represents single en-stabilizing coefficient from VSC *B* to *A* coupled by DC and AC networks. $F_{E13}(s)$ represents single en-stabilizing coefficient from VSC *C* to *A* coupled only by DC network. $F_{E14}(s)$ represents single en-stabilizing coefficient from VSC *C* to *A* coupled by DC and AC networks. $F_{E15}(s)$ represents double en-stabilizing coefficient from VSC *B*, *C* to *A* coupled only by DC network. $F_{E16}(s)$ represents double en-stabilizing coefficient from VSC *B, C* to *A* coupled by DC and AC networks. Detail analytic expressions of these coefficients are calculated in [15], and the results are shown in Appendix *A*.

### *A. Stability index to assess the stability respect to different paths of interactions and parameters*

This part illustrates the effectiveness of the stability index to assessing the stability of the system by Case 1 and 2.

*1) Case 1: Stability index under different* $k_d$ *of I-U droop control*

Assume $F_{E14}(s)$ and $F_{E16}(s)$ are the feedback channels that may have uncertainties, which means $J_0$={1, 2, 3, 5} and $J_1$={4, 6}. Then the transfer function of $G_{1J0}(s)$ and $C_{1J1}(s)$ can be

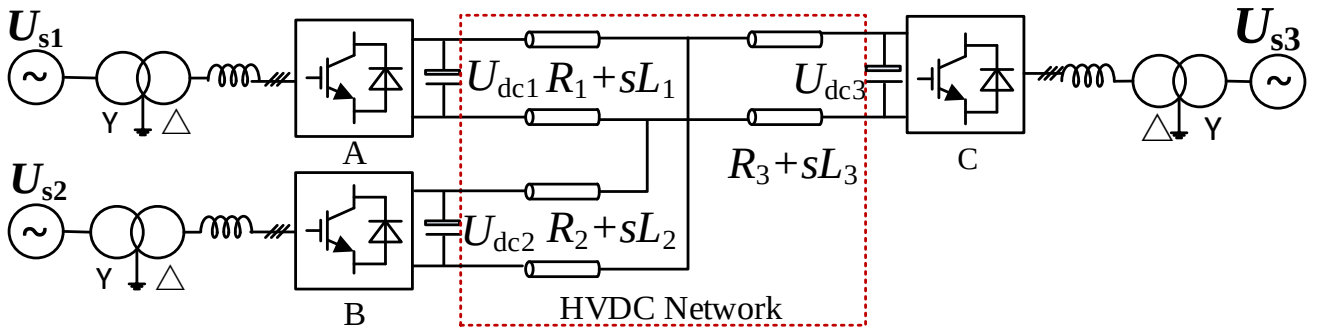

Fig. 6. A typical three-terminal HVDC system.

Table 1. Stability index of the system with different $k_{ad}$ of I-U droop control of VSC *A*

| $k_{ad}$ of I-U droop control of VSC *A* | Stability index $\zeta_{iJ1}$ |
|---|---|
| 1.5 | 0.58 |
| 5 | 0.43 |
| 10 | 0.37 |
| 15 | 0.27 |

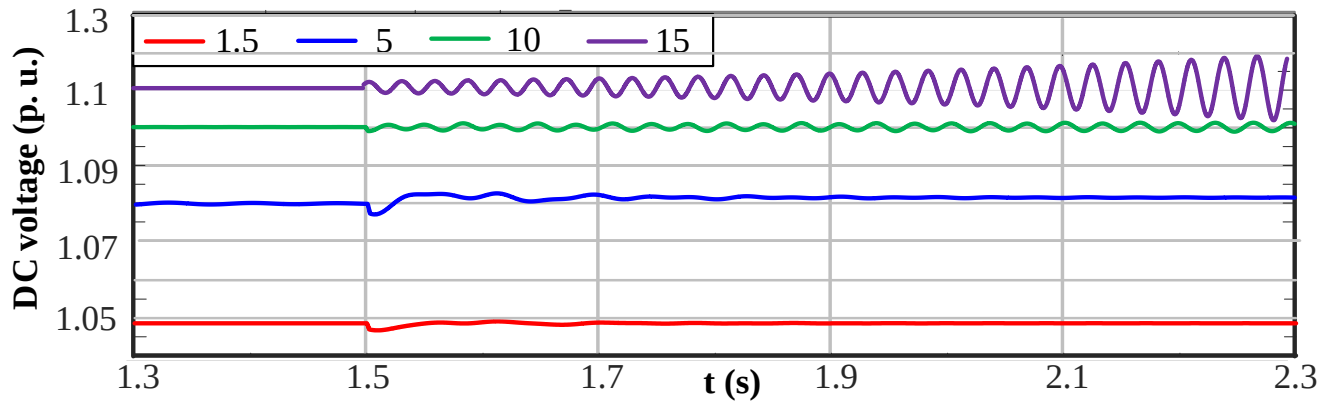

Fig. 7. Time domain responses of DC voltage of VSC A with different $k_d$ of I-U droop control of VSC *A* in Case 1.

Table 2. Stability index of the system respect to different paths of interactions

| Different feedback channels | Stability index $\zeta_{iJ1}$ (rad) |
|---|---|
| $J_1$={4} | 0.51 |
| $J_1$={6} | 0.275 |
| $J_1$={4, 6} | 0.58 |

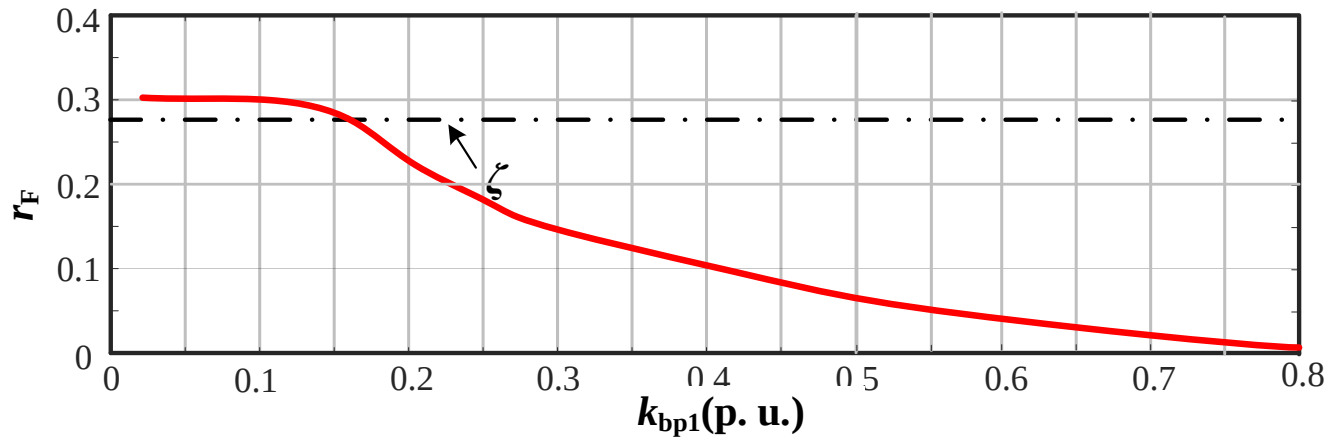

Fig. 8. The maximum value of the ν-gap between - $F_{E16}(s)$ and $-\widetilde{F}_{E16}(s)$ with different with different $k_{bp1}$ in Case 1.

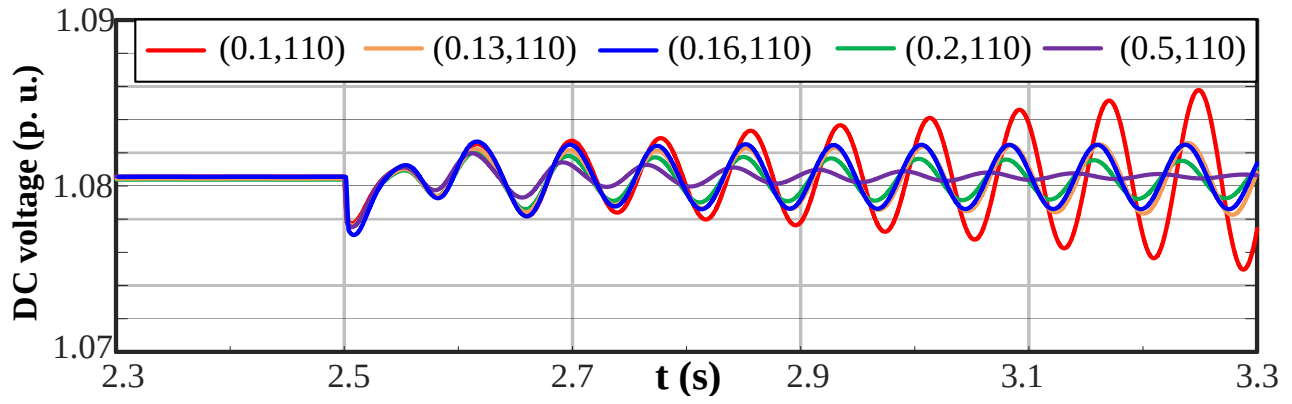

Fig. 9. Time domain responses of DC voltage of VSC *A* with different $k_{bp1}$ in Case 1.

written as

$$G_{1J0} = sC_1U_{dc10} - F_{S1}(s) - F_{E11}(s) - F_{E12}(s) - F_{E13}(s) - F_{E15}(s) \quad (22)$$

$$C_{1J1}(s) = F_{E14}(s) + F_{E16}(s) \quad (23)$$

By substituting(22)-(23) into (14), Table. 1 shows the results of the stability index $\zeta_{1J1}$ with different $k_{ad}$ of I-U droop control of VSC *A*. The operation points and other control parameters are shown in Appendix *C*. Under the same feedforward and feedback channels, the stability index decreases as $k_{ad}$ increases. It means the larger $k_{ad}$, the worse stability performance of the system and the smaller range of uncertainty can be tolerated. The results are verified by time domain simulation depicted in Fig. 7. The system changes from stable to unstable as $k_{ad}$ increases.

2) *Case 2: Stability index respect to different paths of interactions*

Under the set of parameters as shown in Appendix *C*, Table 2 shows the stability index of the system respect to different paths of interactions. The en-stabilizing coefficients with uncertain control parameters, viz. $F_{E12}(s)$, $F_{E14}(s)$ and $F_{E16}(s)$, are mainly considered. Table 2 lists the most common feedback channels and their corresponding stability index. Affected by the topology of DC networks, $F_{E12}(s)$ equals to zero and is ignored due to $A_{12}$=$A_{21}$=0. The results show that the stability index is the smallest when $J_1$={6}, which means the system has the worst stability performance when $F_{E16}(s)$ is the feedback channel. It also shows that the system is 0.275 away from instability respect to this path of interactions. When $J_1$={4, 6}, i.e. the feedback channel is -$F_{E14}(s)$- $F_{E16}(s)$, the system has the best stability performance with the robust stability equals to 0.58.

### *B. The ranges of uncertain parameters preserving the stability*

The following cases are presented to show how the ranges of uncertain control parameters can be evaluate utilizing the proposed method.

1) *Case 3: The range of uncertain proportional parameter of DC Voltage Control in VSC B*

This case calculates the range of the uncertain proportional parameter of DC voltage control in VSC *B*, i.e. $k_{bp1}$. According to the analytic expressions of en-stabilizing coefficients, $k_{bp1}$ only exists in $F_{E12}(s)$ and $F_{E16}(s)$, viz. $J_0$={1, 3, 4, 5} and $J_1$={2, 6}. For $F_{E12}(s)$=0, $k_{bp1}$ influence the stability of the system only through double en-stabilizing coefficient from VSC *B*, *C* to *A* coupled by DC and AC networks. Therefore, the transfer function of feedback channel $C_{1J1}(s)$ equals to $F_{E16}(s)$. The stability index equals to 0.275 according to the results in Table 2.

Calculated by multiple sampling, a curve which depicted the maximum value of the ν-gap between -$F_{E16}(s)$ and $-\widetilde{F}_{E16}(s)$, viz. $r_{1J1}$, with different $k_{bp1}$ is shown Fig. 8. We can see that this curve changes greatly with the change of $k_{bp1}$, indicating that the ν-gap is sensitive to $k_{bp1}$. Moreover, the maximum value of ν-gap between -$F_{E16}(s)$ and $-\widetilde{F}_{E16}(s)$ increases with the decrease of uncertain $k_{bp1}$. In particular, this curve crosses the line representing the stability index $\zeta_{1J1}$ when $k_{bp1}$ is about 0.16, which means the system is in a critically stable state. Thus, $k_{bp1}$ should be designed larger than 0.16 in order to make this MTDC system stable.

2) *Case 4: The range of uncertain proportional parameter of DC Voltage Control in VSC C*

If the proportional parameter of DC voltage control in VSC *C*, i.e. $k_{cp1}$, is uncertain, the range of it is calculated in this case. According to the expressions of en-stabilizing coefficients, $k_{cp1}$ exists in $F_{E14}(s)$ and $F_{E16}(s)$, which means $J_0$={1, 2, 3, 5} and $J_1$={4, 6}. It influences the stability of the system through single en-stabilizing coefficient from VSC *C* to *A* coupled by DC and AC networks, as well as double en-stabilizing coefficient from VSC *B*, *C* to *A* coupled by DC and AC networks. The transfer functions of $G_{1J0}(s)$ and $C_{1J1}(s)$ are written in (22)-(23). The corresponding stability index $\zeta_{1J1}$ equals to 0.58 as shown in Table 2.

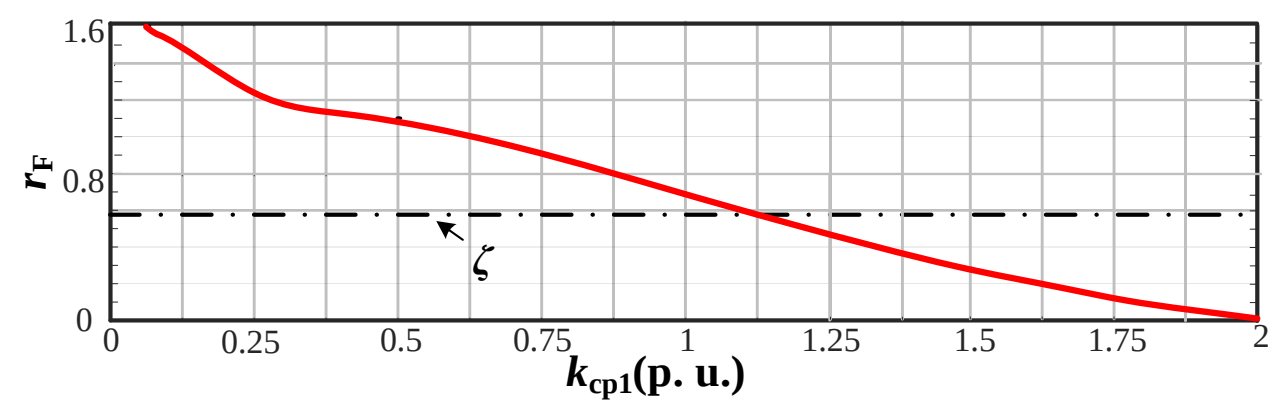

Fig. 10. The maximum value of the ν-gap between -$F_{E14}(s)$-$F_{E16}(s)$ and $-\tilde{F}_{E14}(s)-\tilde{F}_{E16}(s)$ with different with different $k_{cp1}$ in Case 2.

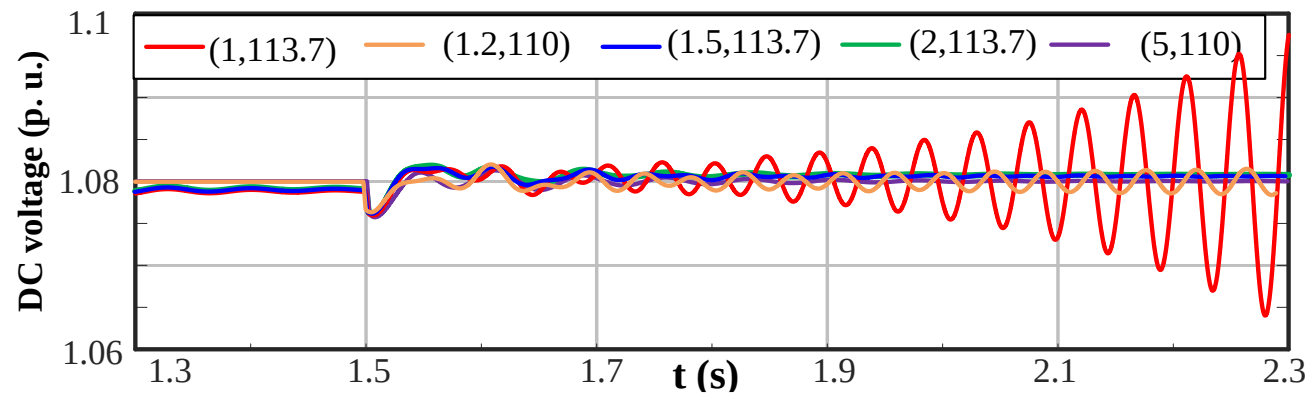

Fig. 11. Time domain responses of DC voltage of VSC *A* with different $k_{cp1}$ in Case 2.

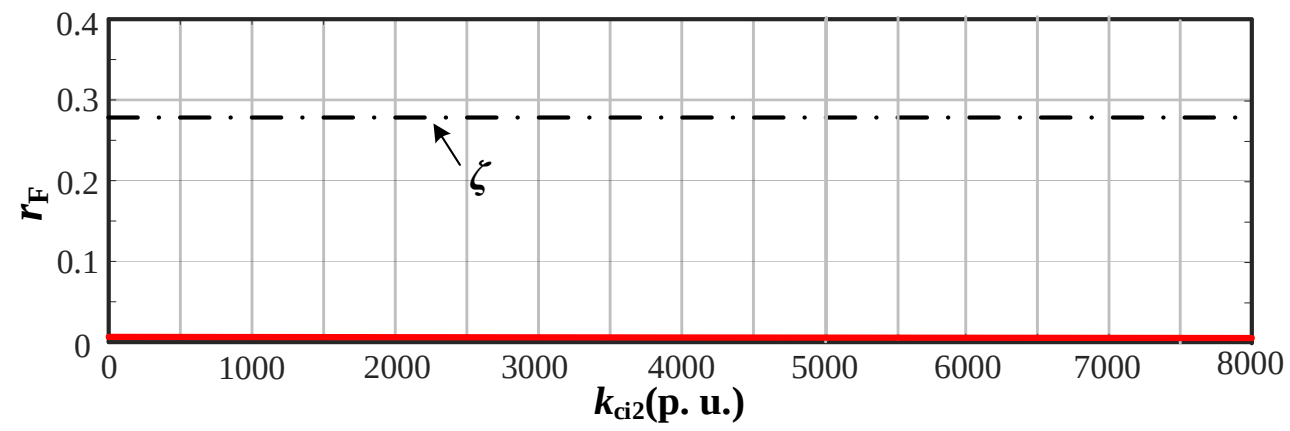

Fig. 12. The maximum value of the ν-gap between -$F_{E14}(s)$-$F_{E16}(s)$ and $-\tilde{F}_{E14}(s)-\tilde{F}_{E16}(s)$ with different $k_{ci2}$ in Case 3.

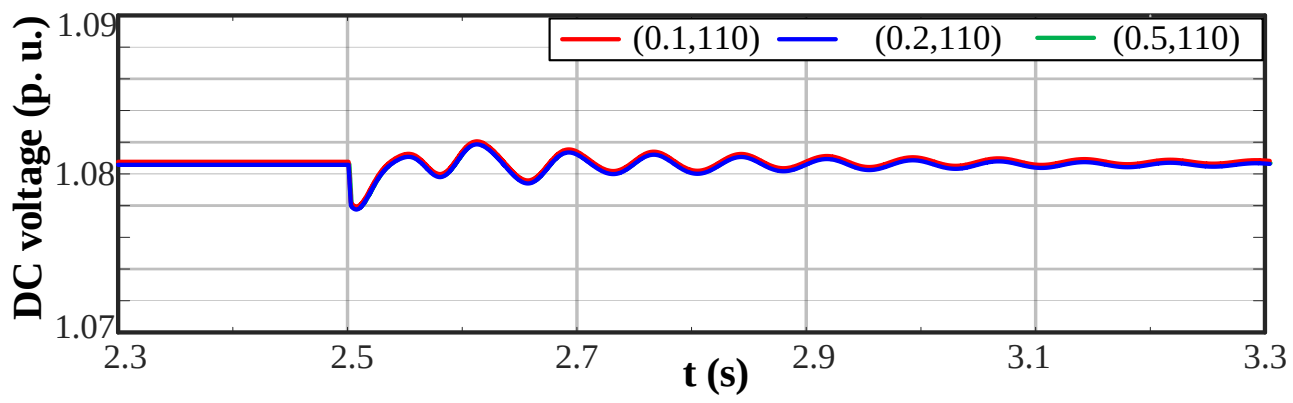

Fig. 13. Time domain responses of DC voltage of VSC A with different $k_{ci2}$ in Case 3.

Fig. 10 shows how the maximum value of the ν-gap between -$F_{E14}(s)$-$F_{E16}(s)$ and $-\tilde{F}_{E14}(s)-\tilde{F}_{E16}(s)$, viz. $r_{1J1}$, changes with respect to different $k_{cp1}$. The value of $r_{1J1}$ changes significantly while the variation of $k_{cp1}$ is small, indicating that the ν-gap is sensitive to $k_{cp1}$. And $r_{1J1}$ increases with the decrease of $k_{cp1}$. Furthermore, the system is critically stable when this curve crosses the stability index $\zeta_{1J1}$ at about $k_{cp1}$ =1.125. Therefore, $k_{cp1}$ should be designed larger than 1.125 to make the system stable.

3) *Case 5: Effect of Uncertain integral parameter of PLL in VSC C*

This case calculates the range of the uncertain integral parameter of PLL in VSC *C*, viz. $k_{ci2}$. According to the expressions of en-stabilizing coefficients, $k_{ci2}$ exists in $F_{E14}(s)$ and $F_{E16}(s)$ as the same as *Case* 4. When $k_{ci2}$ is 2000, the corresponding stability index $\zeta_{1J1}$ is 0.58, as same as *Case* 4.

Fig. 12 shows the frequency response of the ν-gap $\nu\left[-\tilde{F}_{E14}(s)-\tilde{F}_{E16}(s),-F_{E14}(s)-F_{E16}(s)\right]$ with different $k_{ci2}$. It can be seen that the maximum value of the ν-gap $r_{1J1}$ is close to zero, and the curve changes little with the change of $k_{ci2}$ indicating that the ν-gap is not sensitive to $k_{ci2}$. The results prove mathematically that the uncertain integral parameter of PLL has little effect on the system stability under strong AC network conditions. The time domain simulation in Fig. 13 verifies the analysis.

## V. Conclusions

The stability index of an MTDC system is defined to characterize the relative stability of a particular VSC with respect to different paths of interactions with uncertainty. Generally, the larger the stability index, the better the stability performance of the MTDC system. The ν-gap metric is applied to measure the distance of the nominal system of interactions and the system of uncertain interactions. Based on the difference between the stability index and the ν-gap metric, we have presented the necessary and sufficient condition of the system being stable. By this condition, we can quantitatively know how far the system is away from instability.

A method for calculating the range of uncertain parameters preserving the stability has been proposed. We can also analyze the sensitivity of a stable MTDC system to a particular parameter by this method based on the variation of the ν-gap between certain and uncertain systems with respect to the variation of the parameter.

In summary, the results of this paper can explain the influence of control parameters on robust stability through interactions among VSCs quantitatively, and provide new ideas for multi-equipment-based stability control design.

## Appendix

### *A. Analytic expressions of different paths of self-/en-stabilizing coefficients*

$$F_{E11}(s)=\frac{A_{12}A_{21}\left(K_{P01}G_{in1}(s)+1\right)}{sC_2U_{dc20}-A_{22}}$$

$$F_{E12}(s)=\left[\frac{A_{12}A_{21}\left(G_{in1}(s)K_{P01}+1\right)}{sC_1U_{dc10}+G_{dc1}(s)K_{P01}-A_{22}G_{in1}(s)K_{P01}-A_{22}}-\frac{A_{12}A_{21}}{sC_1U_{dc10}-A_{22}}\right]\bullet\left(K_{P01}G_{in1}(s)+1\right)$$

$$F_{E13}(s)=\frac{A_{13}A_{31}\left(K_{P01}G_{in1}(s)+1\right)}{sC_3U_{dc30}-A_{33}}$$

$$F_{E14}(s)=\left[\frac{A_{13}A_{31}\left(G_{in3}(s)K_{P03}+1\right)}{sC_3U_{dc30}+G_{dc3}(s)K_{P03}-A_{33}G_{in3}(s)K_{P03}-A_{33}}-\frac{A_{13}A_{31}}{sC_3U_{dc30}-A_{33}}\right]\bullet\left(K_{P01}G_{in1}(s)+1\right)$$

$$
\begin{aligned}
F_{\mathrm{E15}}(s)=&[\left(A_{12}\left(A_{21}C_3U_{\mathrm{dc30}}s-A_{21}A_{33}+A_{31}A_{23}\right)\right)\\
&/(C_2C_3U_{\mathrm{dc20}}U_{\mathrm{dc30}}s^2-A_{22}C_3U_{\mathrm{dc30}}s-A_{33}C_2U_{\mathrm{dc20}}s\\
&+A_{22}A_{33}-A_{23}A_{32})\\
&+\left(A_{13}\left(A_{31}C_2U_{\mathrm{dc20}}s+A_{21}A_{32}-A_{22}A_{31}\right)\right)\\
&/(C_2C_3U_{\mathrm{dc20}}U_{\mathrm{dc30}}s^2-A_{22}C_3U_{\mathrm{dc30}}s-A_{33}C_2U_{\mathrm{dc20}}s\\
&+A_{22}A_{33}-A_{23}A_{32})-A_{12}A_{21}/(sC_2U_{\mathrm{dc20}}-A_{22})\\
&-A_{13}A_{31}/(sC_3U_{\mathrm{dc30}}-A_{33})]\bullet\left(K_{\mathrm{P\theta1}}G_{\mathrm{in1}}(s)+1\right)
\end{aligned}
$$

$$
\begin{aligned}
F_{\mathrm{E16}}(s)=&\left[(-A_{21}A_{33}A_{12}+A_{12}A_{23}A_{31}+A_{13}A_{21}A_{32}\right.\\
&-A_{13}A_{22}A_{31})\left(K_{\mathrm{P\theta3}}G_{\mathrm{in3}}(s)+1\right)\left(K_{\mathrm{P\theta2}}G_{\mathrm{in2}}(s)+1\right)\\
&+A_{12}A_{21}\left(K_{\mathrm{P\theta2}}G_{\mathrm{in2}}(s)+1\right)\left(K_{\mathrm{P\theta3}}G_{\mathrm{dc3}}(s)+sC_3U_{\mathrm{dc30}}\right)\\
&\left.+A_{13}A_{31}\left(K_{\mathrm{P\theta3}}G_{\mathrm{in3}}(s)+1\right)\left(K_{\mathrm{P\theta2}}G_{\mathrm{dc2}}(s)+sC_2U_{\mathrm{dc20}}\right)\right]\\
&/\left[(A_{22}A_{33}-A_{23}A_{32})\left(K_{\mathrm{P\theta3}}G_{\mathrm{in3}}(s)+1\right)\left(K_{\mathrm{P\theta2}}G_{\mathrm{in2}}(s)+1\right)\right.\\
&-A_{22}\left(K_{\mathrm{P\theta3}}G_{\mathrm{dc3}}(s)+sC_3U_{\mathrm{dc30}}\right)\left(K_{\mathrm{P\theta2}}G_{\mathrm{in2}}(s)+1\right)\\
&-A_{33}\left(K_{\mathrm{P\theta2}}G_{\mathrm{\theta U2}}(s)+sC_2U_{\mathrm{dc20}}\right)\left(K_{\mathrm{P\theta3}}G_{\mathrm{in3}}(s)+1\right)\\
&\left.+\left(K_{\mathrm{P\theta3}}G_{\mathrm{dc3}}(s)+sC_3U_{\mathrm{dc30}}\right)\left(K_{\mathrm{P\theta2}}G_{\mathrm{dc2}}(s)+sC_2U_{\mathrm{dc20}}\right)\right]\\
&\bullet\left(K_{\mathrm{P\theta1}}G_{\mathrm{in1}}(s)+1\right)-F_{\mathrm{E11}}(s)-F_{\mathrm{E12}}(s)-F_{\mathrm{E13}}(s)\\
&-F_{\mathrm{E14}}(s)-F_{\mathrm{E15}}(s)
\end{aligned}
$$

## *B. System parameters for cases 1~3*

1) Steady state reference values for Case 1~3
   Inspecting power $S_{\mathrm{base}}$=1000MVA
   Line voltage effective value $U_{\mathrm{base}}$=270kV
   Frequency $f_{\mathrm{base}}$=50Hz
   DC voltage $U_{\mathrm{dcbase}}$=400kV
2) Per-unit values for Case 1~3

*VSC A* $P_{10}$=-0.5 p.u., $Q_{10}$= 0 p.u.
$U_{\mathrm{dc10}}$ =1 p.u., $U_{\mathrm{t10}}$ =1 p.u., $U_{\mathrm{s10}}$ =1 p.u.
$C_1$=0.05 p.u., $X_{\mathrm{f1}}$=0.25 p.u., $X_{\mathrm{g1}}$=0.2 p.u.
$R_1$=0.4 p.u., $L_1$=0.01 p.u.

*VSC B* $P_{20}$=-0.3 p.u., $Q_{20}$= 0 p.u.
$U_{\mathrm{dc20}}$ =1 p.u., $U_{\mathrm{t20}}$ =1 p.u., $U_{\mathrm{s20}}$ =1 p.u.
$C_2$=0.03 p.u., $X_{\mathrm{f2}}$=0.25 p.u., $X_{\mathrm{g2}}$=0.2 p.u.
$R_2$=0.4 p.u., $L_2$=0.01 p.u.

*VSC C* $P_{30}$=0.8 p.u., $Q_{30}$= 0 p.u.
$U_{\mathrm{dc30}}$ =1 p.u., $U_{\mathrm{t30}}$ =1 p.u., $U_{\mathrm{s30}}$ =1 p.u.
$C_3$=0.05 p.u., $X_{\mathrm{f3}}$=0.25 p.u., $X_{\mathrm{g3}}$=0.2 p.u.
$R_3$=0 p.u., $L_3$=0 p.u.

## *C. Controller Parameters in case 1~3 without uncertainties*

| | | |
|---|---|---|
| PLL control | $k_{\mathrm{ap2}}$ =25 | $k_{\mathrm{ai2}}$ =2000 |
| | $k_{\mathrm{bp2}}$ =50 | $k_{\mathrm{bi2}}$ =1000 |
| | $k_{\mathrm{cp2}}$ =50 | $k_{\mathrm{ci2}}$ =2000 |
| Droop control | $k_{\mathrm{ad}}$=$k_{\mathrm{bd}}$=$k_{\mathrm{cd}}$=1.5 | |
| DVC control | $k_{\mathrm{ap1}}$ =1.3 | $k_{\mathrm{ai1}}$ =113.7 |
| | $k_{\mathrm{bp1}}$ =0.8 | $k_{\mathrm{ai1}}$ =110 |
| | $k_{\mathrm{cp1}}$ =2 | $k_{\mathrm{ci1}}$ =113.7 |
| AC current control | $k_{\mathrm{ap6}}$ = $k_{\mathrm{bp6}}$ = $k_{\mathrm{cp6}}$ =0.6 | |
| | $k_{\mathrm{ai6}}$ =$k_{\mathrm{bi6}}$ = $k_{\mathrm{ci6}}$ =640 | |

## REFERENCES

[1] Gunnar Asplund, Kerstin Lindén, Carl Barker, "HVDC grid feasibility study," Cigré Working Group B4.52, Apr. 2013

[2] V. Akhmatov, M. Callavik, and C.M. Franck, "Technical guidelines and prestandardization work for first HVDC grids," *IEEE Trans. Power Del.*, vol. 29, no. 4, pp. 327-335, Feb. 2014.

[3] G. O. Kalcon, G. P. Adam, and O. A. Lara, "Small-signal stability analysis of multi-terminal VSC-based DC transmission systems," *IEEE Trans. Power Syst.*, vol. 27, no. 4, pp. 1818-1830, Nov. 2012.

[4] S. Cole, J. Beerten, and R. Belmans, "Generalized dynamic VSC MTDC model for power system stability studies," *IEEE Trans. Power Syst.*, vol. 25, no. 3, pp. 1655-1662, Aug. 2010.

[5] N. R. Chaudhuri, R. Majumder, and B. Chaudhuri, "Stability analysis of VSC MTDC grids connected to multimachine AC systems," *IEEE Trans. Power Del.*, vol. 26, no. 4, pp. 2774-2784, Oct. 2011.

[6] G. Stamatiou, "Converter interactions in VSC-based HVDC systems," *Lic. of Eng. Thesis, Dept. Energy Environ.*, Chalmers Univ. Technol., Gothenburg, Sweden, Jun. 2015.

[7] W. Du, Q. Fu, and H. Wang, "Small-signal stability of an AC/MTDC power system as affected by open-loop modal coupling between the VSCs," *IEEE Trans. Power syst.*, vol. 33, no.3, pp. 3143-3152, May. 2018.

[8] Z, Zhen, W. Du and H. Wang, "The proof of the equivalence of complex torque coefficient method and open-loop mode stabilization judgement method in strong mode coupling phenomena," in *Proceedings of the CSEE*, vol. 39, no. 8, pp. 2272-2279, Apr. 2019.

[9] Y. Zhan, X. Xie, H. Liu, "Frequency-domain modal analysis of the oscillatory stability of power systems with high-penetration renewables," *IEEE Trans. on Sustain. Energy*, vol. 10, no. 3, pp. 1534-1543, Jul. 2019.

[10] M. Cespedes, J. Sun, "Impedance modeling and analysis of grid-connected voltage-source converters," *IEEE Trans. Power Electron.*, vol. 29, no. 3, pp. 1254-1261, Mar. 2014.

[11] J. Sun and H. Liu, "Sequence impedance modeling of Modular Multilevel Converters," *IEEE J. Emerg. Sel. Topics Power Electron.*, vol. 5, no. 4, pp. 1427-1443, Dec. 2017.

[12] H. Liu and J. Sun, "Impedance-based stability analysis of VSC-based HVDC systems," in *Proc. 2013 IEEE COMPEL*, Salt Lake City, USA, Jun. 2013.

[13] L. Xu, L. Fan, and Z. Miao, "DC impedance-model-based resonance analysis of a VSC–HVDC system," *IEEE Trans. Power Del.*, vol. 30, no. 3, pp.1221-1230, Jun. 2015.

[14] J. Lyu, X. Cai, and M. Molinas, "Frequency domain stability analysis of MMC-based HVDC for wind farm integration" *IEEE J. Emerg. Sel. Topics Power Electron.*, vol. 4, no. 1, pp.141-151, Mar. 2016.

[15] W. Zheng, J. Hu and X. Yuan, "Quantification of Interactions in MTDC Systems based on Self-/En-Stabilizing Coefficients in DC Voltage Control Timescale," *IEEE J. Emerg. Sel. Topics Power Electron.*, vol. 9, no. 3, pp. 2980-2991, Jun. 2021.

[16] L. Huang, H. Xin and F. Dörfler, "H∞-control of grid-connected converters: design, objectives and decentralized stability certificates," *IEEE Tran. on Smart Grid*, vol. 11, no.5, pp. Sep. 2020.

[17] S. Yang, Q. Lei, and F. Z. Peng, "A robust control scheme for grid-connected voltage-source inverters," *IEEE Trans. Ind. Electron.*, vol. 58, no. 1, pp. 202–212, Jan. 2011.

[18] C. Kammer, S. D'Arco, and A. G. Endegnanew, "Convex optimization-based control design for parallel grid-connected inverters," *IEEE Trans. Power Electron.*, vol. 34, no. 7, pp. 6048–6061, Jul. 2019.

[19] E. Sánchez-Sánchez, D. Groß, E. Prieto-Araujo, F. Dörfler, and O. Gomis-Bellmunt, "Optimal multivariable MMC energy-based control for dc voltage regulation in HVDC applications," *IEEE Trans. Power Del.*, vol. 35, no. 2, pp. 999–1009, Apr. 2020.

[20] G. Vinnicombe, Uncertainty and Feedback, H∞ Loop-shaping and the ν-gap Metric, World Scientific Pub Co., 1999.

[21] L. Qiu and K. Zhou, Introduction to Feedback Control, Prentice Hall, 2009.

[22] X. Yuan, S. Cheng, and J. Hu, "Multi-time scale voltage and power angle dynamics in power electronics dominated large power systems," in *Proceedings of CSEE.*, vol.36, no.19, pp. 5145-5154, Oct, 2016.

[23] W. Zheng, J. Hu and X. Yuan, "Modeling of MTDC-linked VSCs considering input and output active power dynamics for multi-terminal HVDC interaction analysis in DC Voltage Control Timescale," *IEEE Trans. Energy Convers.*, vol. 34, no. 4, pp. 2008-2018, Dec. 2019.

[24] W. Zheng, J. Hu and K. Liu, "Interaction analysis of MTDC systems based on self-/en-stabilizing coefficients in weak AC grid conditions in DC voltage control timescale," in *2021 IEEE ECCE-Asia*, Singapore, May, 2021.